\documentclass[aps,pre,twocolumn,groupedaddress,longbibliography]{revtex4-1}
\usepackage{graphicx,amsmath,amssymb,color,subfigure}

\begin{document}

\author{Pedro Abritta}
\author{Robert S. Hoy}
\affiliation{Department of Physics, University of South Florida, Tampa, FL 33620 USA}
\email{rshoy@usf.edu}
\date{\today}

\title{Structure of saturated RSA ellipse packings}

\begin{abstract}
Motivated by the recent observation of liquid glass in suspensions of ellipsoidal colloids, we examine the structure of (asymptotically) saturated RSA ellipse packings. 
We determine the packing fractions $\phi_{\rm s}(\alpha)$ to high precision, finding an { empirical} analytic { formula} that predicts $\phi_{\rm s}(\alpha)$ to within less than $0.1\%$ for all $\alpha \leq 10$.
Then we explore how these packings' positional-orientational order varies with $\alpha$.
We find a transition from tip/side- to side/side-contact-dominated structure at $\alpha = \alpha_{\rm TS} \simeq 2.4$.
At this aspect ratio, the peak value $g_{\rm max}$ of packings' positional-orientational pair correlation functions is minimal, and systems can be considered maximally { locally} disordered.
For smaller (larger) $\alpha$, $g_{\rm max}$ increases exponentially with deceasing (increasing) $\alpha$.   
Local nematic order and structures comparable to the precursor domains observed in experiments gradually emerge as $\alpha$ increases beyond $3$.
For $\alpha \gtrsim 5$, single-layer lamellae become more prominent, and long-wavelength density fluctuations increase with $\alpha$ as packings gradually approach the rod-like limit.
\end{abstract}

\maketitle

\section{Introduction}
\label{sec:intro}

Recent advances in colloidal synthesis and microscopy techniques have dramatically improved our ability to characterize how particle ordering and relaxation in thermal liquids varies with particle shape.
For example, high-quality monodisperse colloidal rods and cylinders can now be produced \cite{dugyala13}, and their liquid-state dynamics (in suspensions) can be observed using confocal microscopy and allied techniques \cite{weeks00,gasser01}.
Of particular interest is the recent experimental observation of {``}liquid glass{''}.
Roller \textit{et al.}\ found \cite{roller21} that suspensions of ellipsoidal colloids with aspect ratio $\alpha_{\rm R} = 3.5$ exhibited two distinct glass transitions at packing fractions $\phi_{\rm g}^{\rm rot}$ and $\phi_{\rm g}^{\rm trans}$.
In the liquid glass state ($\phi_{\rm g}^{\rm rot}  <  \phi < \phi_{\rm g}^{\rm trans}$), particles rotations' are arrested, but they remain free to translate within  locally-nematic precursor domains.
This phenomenon had been predicted by simulations \cite{letz00}, but had previously only been experimentally observed in quasi-2D systems \cite{zheng11,zheng14,mishra13}. 
Experiments like Ref.\ \cite{roller21} offer both a new avenue for understanding the physics of anisotropic molecular glassformers and an obvious motivation for theoretical studies of related models.

For hard ellipses and ellipsoids of revolution with aspect ratio $\alpha$, complex liquid-state dynamics are expected for packing fractions in the range $\phi_{\rm {o}}(\alpha) < \phi < \phi_{\rm g}(\alpha)$, where $\phi_{\rm {o}}(\alpha)$ is the ``onset'' density \cite{sastry98,chaudhuri10} and $\phi_g(\alpha)$ is either the rotational or translational glass transition density.
Evaluating $\phi_{\rm {o}}(\alpha)$ and $\phi_{\rm g}(\alpha)$ using simulations is very { computationally expensive} \cite{pfleiderer08,shen12,davatolhagh12,xu15}.
A more easily obtained yet physically relevant packing fraction that lies between $\phi_{\rm {o}}(\alpha)$ and $\phi_g(\alpha)$ is the random sequential adsorption (RSA) density $\phi_{\rm s}(\alpha)$, the maximum density at which impenetrable ellipses of aspect ratio $\alpha$ can be packed under a protocol that sequentially inserts them with random positions and orientations.
The differences $\phi_{\rm g}(\alpha) - \phi_{\rm s}(\alpha)$ and $\phi_{\rm J}(\alpha) - \phi_{\rm s}(\alpha)$ are of particular interest because they indicate how much packing efficiency can be gained (before glass formation and jamming, respectively) by allowing particles to move freely while remaining positionally and orientationally disordered.  
For example, in the $\alpha \to \infty$ limit one expects $\phi_{\rm J}(\alpha) = \phi_{\rm s}(\alpha)$ because particle rotations are completely blocked \cite{philipse96,desmond06}.

While jamming of ellipses and ellipsoids is now fairly well understood \cite{donev04,donev07,delaney05,schreck09}, RSA of these systems remains relatively poorly characterized.
$\phi_{\rm s}(\alpha)$ values for ellipses have been reported only for $1 \leq \alpha \leq 5$ \cite{sherwood90,viot92,ciesla16,haiduk18}, and detailed characterization of RSA ellipse packings' structure has only been performed for the aspect ratio $\alpha \simeq 1.85$ that maximizes $\phi_{\rm s}$ \cite{ciesla16}.
Thus there is a need to substantially expand our knowledge of these packings.

In this paper, we characterize saturated RSA ellipse packings over a wider range of aspect ratios and in much greater detail than has been previously attempted.
First we determine their packing fractions $\phi_{\rm s}(\alpha)$ to within $\sim 0.1\%$ for $1 \leq \alpha \leq 10$.
Then we characterize their positional-orientational order using several metrics.
We find a previously-unreported structural transition at $\alpha = \alpha_{\rm TS} \simeq 2.4$.
For $1 < \alpha < \alpha_{\rm TS}$ ($\alpha > \alpha_{\rm TS}$), packings have an excess of tip-to-side (side-to-side) contacts.
The peak prevalence of the favored type of contact, an effective order parameter for these systems, increases exponentially with $|\alpha - \alpha_{\rm TS}|$. 
We also show that (i) local nematic order and structures comparable to the precursor domains observed in experiments \cite{roller21,zheng11} gradually emerge as $\alpha$ increases beyond $3$, and (ii) the increasing size of single-layer lamellae that are randomly oriented within these packings makes their long-wavelength density fluctuations increase rapidly with $\alpha$ for $\alpha \gtrsim 5$.

\section{Generating Saturated Packings}

Saturated RSA packings of anistropic 2D particles are generated by placing them with random positions and orientations, typically within square domains of size $L \times L$, until no more particles can be inserted.
In practice,  RSA packing generation's inherently slow kinetics \cite{swendsen81} make achieving complete saturation for $L$ that are large enough to minimize finite-size effects extremely difficult.
Therefore we employ an efficient protocol that generates packings that are demonstrably within $< 1\%$ (and for most $\alpha$, much less than $1\%$) of their saturation densities, for system sizes that are in the $L \to \infty$ limit.
Specifically, we use $L = 1000{\sigma}${, where $\sigma$ is the length of the ellipses' minor axes.
This} is the same {$L$} employed in Refs.\ \cite{ciesla16,haiduk18} and is sufficiently large that finite-size errors on $\phi_{\rm s}$ should be less than $0.1\%$ \cite{ciesla16}.
{ Below, we set $\sigma = 1$.}

We divide our periodic domains into $\Lambda = \rm{floor}(L/\alpha)^2$ linked cells of size $(L/\Lambda) \times (L/\Lambda)$.
Here $\rm{floor}[x]$ rounds $x$ downward to the nearest integer, e.g.\ $\rm{floor}[4.6] = 4$.
Thus, when insertion of an ellipse $i$ at position $\vec{r}_i$ is attempted, only ellipses in the cell containing $\vec{r}_i$ 
and its 8 neighboring cells need be checked for overlap.
Checks for overlap with the set $\{ j \}$ of all ellipses that have already been inserted in these cells (at positions $\{ \vec{r}_j \}$) include three steps: 
(1) Since we assume ellipses' minor axes have unit length while their major axes have length $\alpha$, any center-to-center distances $r_{ij} = |\vec{r}_j - \vec{r}_i| < 1$ imply overlap, and the insertion attempt is rejected. 
(2) If $r_{ij} > \alpha$, the ellipses do not overlap and the code continues to the next $j$. 
(3) If $1 \leq r_{ij} \leq \alpha$, overlap is possible. 
 We determine whether the ellipses overlap using  Zheng and Palffy-Muhoray's exact expression \cite{zheng07} for their distance of closest approach $d_{\rm cap}$.
{ For monodisperse ellipses with unit-length minor axes, their expression reduces to 
\begin{equation}
d_{cap} = \frac{ d' }{\sqrt{ 1 -  (1 - \alpha^{-2})(\hat{k_i}\cdot\hat{r}_{ij})^2 } },
\label{eq:dcap}
\end{equation}
where $d'$ is obtained using a complicated formula involving $r_{ij}$ and the ellipses' orientation vectors $\hat{k}_i$ and $\hat{k}_j$.
Note that $\hat{k}_j$ does not appear in Eq.\ \ref{eq:dcap} because it has been absorbed into $d'$; cf.\ Eqs.\ 33, 36 of Ref.\ \cite{zheng07}. }
If $d_{cap} > r_{ij}$, the ellipses overlap and the insertion attempt is rejected.
Otherwise the ellipse is inserted and the next insertion attempt begins.

We examined 81 different particle aspect ratios over the range $1 \leq \alpha \leq 10$.
To allow extrapolation of the runs' progress to the infinite-time limit, we tracked the packing fractions $\phi(\alpha,t)$ after $t$ insertion attempts per unit area had been completed.
Each run continued until $2.5\times10^5$ trials per unit area had been attempted; this was sufficient to reach well into the asymptotic $t^{-1/3}$ regime \cite{swendsen81} for all $\alpha$.
Then we determined the RSA densities $\phi_{\rm s}(\alpha)$ by fitting the { results to}
\begin{equation}
\phi(\alpha,t) = \phi_{\rm s}(\alpha) \left(1 - \left[ \frac{t}{\tau(\alpha)}\right]^{-1/3} \right) {,}
\label{eq:t13}
\end{equation}
{ where $\tau(\alpha)$ is a ``time'' characterizing the $\alpha$-dependent RSA kinetics.}

\begin{figure}[h]
\includegraphics[width=3in]{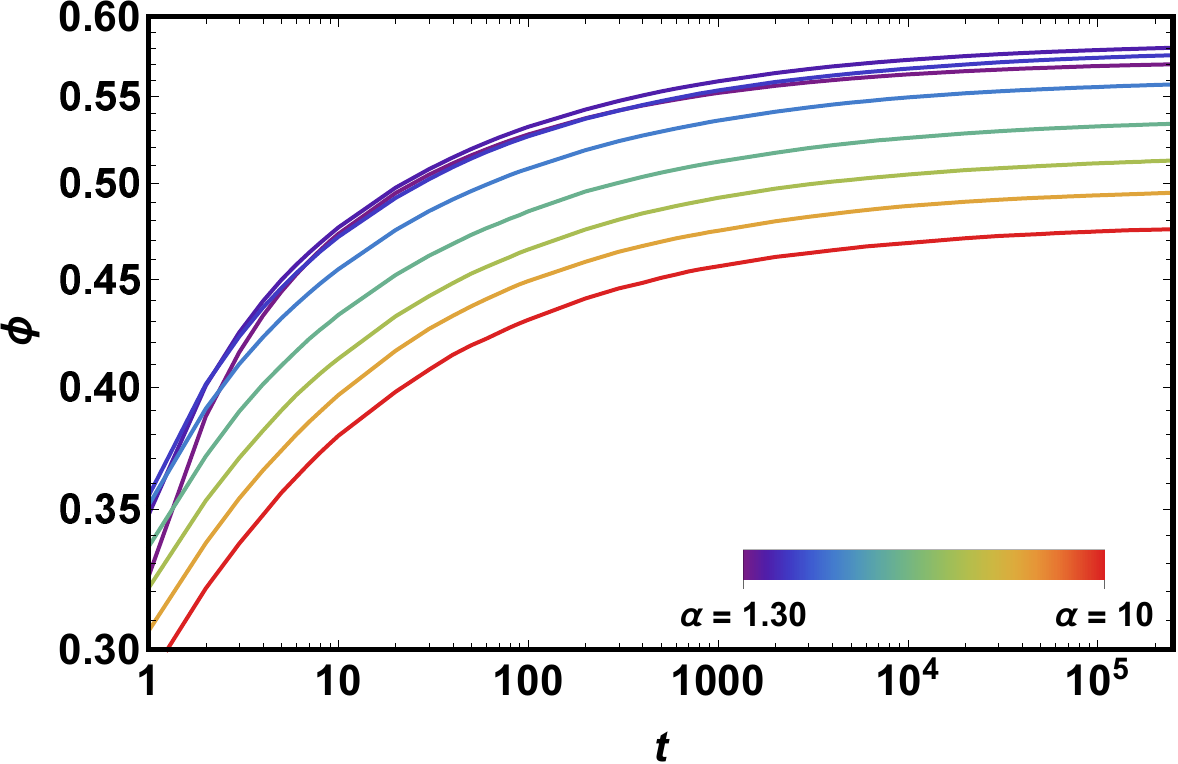}
\caption{RSA kinetics for ellipses.  Results are shown for  $\alpha = 1.3$, $1.85$, $2.4$, $3.5$, $5$, $6.5$, $8$, and $10$.}
\label{fig:phivst}
\end{figure}

Figure \ref{fig:phivst} shows {$\phi(\alpha,t)$} 
for selected $\alpha$.
There is some crossing of the curves at intermediate $t$ for $\alpha < 2$ because $\phi_{\rm s}(\alpha)$ is non-monotonic, but as expected for anisotropic particles \cite{vigil89}, $\tau$ increases with $\alpha$.
All $\alpha \leq 10$ had $\tau < 0.13$ and hence had $\phi_{\rm s}(\alpha)/\phi(\alpha, 2.5\times10^5) - 1 < .0082$.
In other words, at {the end of our packing-generation runs,}
all $\alpha \leq 10$ have $\phi$ that are within less than $1\%$ of $\phi_{\rm s}(\alpha)$.

\section{Structure of Saturated Packings}

\begin{figure}[h!]
\includegraphics[width=3.25in]{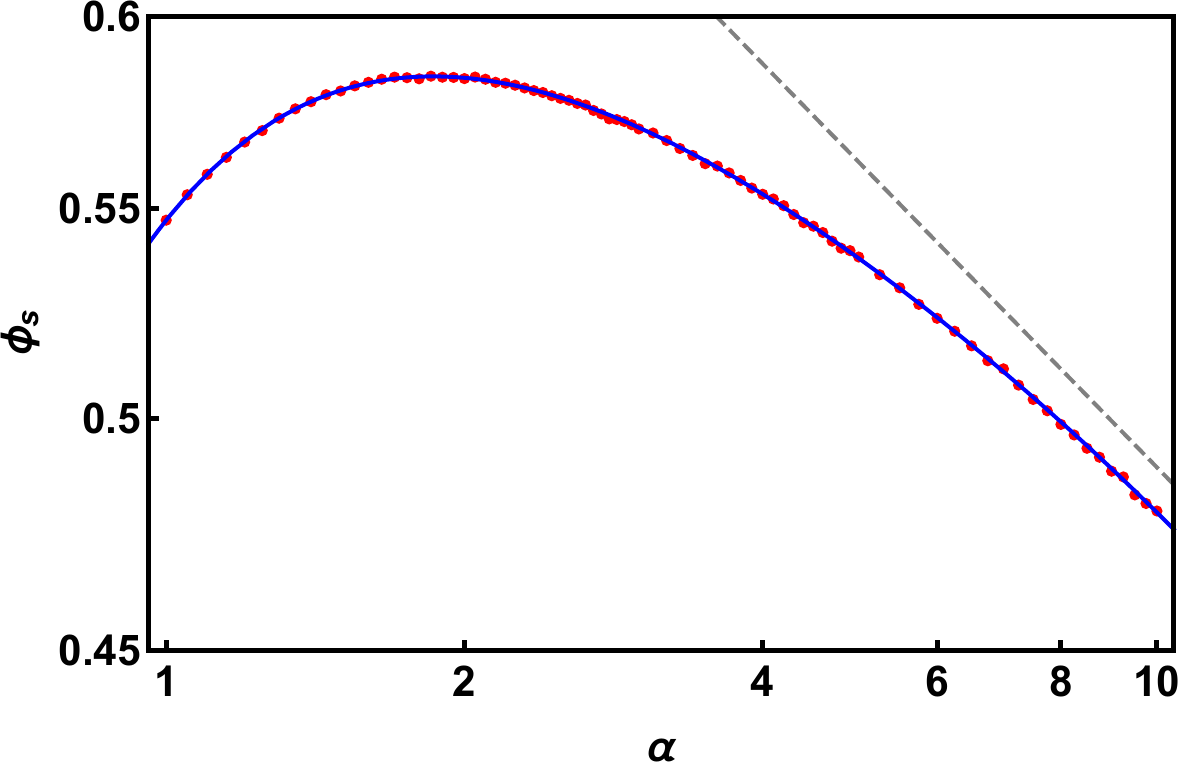}
\caption{Packing fractions of saturated {RSA} ellipse packings..
Red circles show the $\phi_{\rm s}(\alpha)$ estimated from Eq.\ \ref{eq:t13}, while the blue curve shows Eq.\ \ref{eq:analyticphis}.  The dashed gray line shows $\phi = 0.775\alpha^{-1/5}$ and is included only to illustrate the lack of convergence to $\alpha^{-1}$ scaling.}
\label{fig:phis}
\end{figure}

Figure \ref{fig:phis} shows the extrapolated $\phi_{\rm s}(\alpha)$.  
Results for all $\alpha \leq 5$ agree with previous studies.
In particular, they are consistent with the well-known disk RSA packing fraction $\phi_{\rm s}(1) = \phi_{\rm disks} = .54707$ \cite{feder80,zhang13} and the known density maximum at $\alpha \simeq 1.85$, i.e.\ $\phi_{\rm s}(1.85) \simeq .584$ \cite{sherwood90,viot92,ciesla16,haiduk18}.
For large aspect ratios, our results show a surprisingly slow decrease of $\phi_s$ with increasing $\alpha$.
Specifically, since  the area swept out by ellipses as they rotate about their centers is $A_{\rm sw} = \pi\alpha^2/4$ whereas the area of the ellipses themselves is $\pi \alpha/4$, Onsager's classical arguments \cite{onsager49} predict $\phi_{\rm s} \sim A/A_{\rm sw} = 1/\alpha$.
Our data show that this asymptotic regime is not reached until $\alpha \gg 10$.

We find that  the extrapolated $\phi_{\rm s}(\alpha)$ are well fit by 
\begin{equation}
\phi_{\rm s}(\alpha) = \phi_{\rm disks} \times \displaystyle\frac{1 + (3/8)\ln(\alpha) +  (17/25)(\alpha-1)}{1 + (80/99)(\alpha - 1) + (\alpha-1)^2/96}.
\label{eq:analyticphis}
\end{equation}
$\phi_{\rm s}(\alpha) > \phi_{\rm disks}$ for all $1 < \alpha \lesssim 4.40$, indicating that particles' anisotropy enhances their packability over this range of aspect ratios.
Equation \ref{eq:analyticphis} is an empirical formula designed to capture the known trends: convergence to $\phi_{\rm disks}$ as $\alpha \to 1$, and $1/\alpha$ scaling at large $\alpha$ {\cite{philipse96,desmond06}}.
Its mean and rms fractional deviations from the extrapolated $\phi_{\rm s}(\alpha)$ are respectively $0.02\%$ and $0.06\%$.
We do \textit{not} claim that Eq.\ \ref{eq:analyticphis} is an exact expression valid for all $\alpha$,
{ or even that its functional form is the same as that of the ``true'' $\phi_s(\alpha)$ which could be obtained given infinite computer power.}

\begin{figure}[h!]
\includegraphics[width=3in]{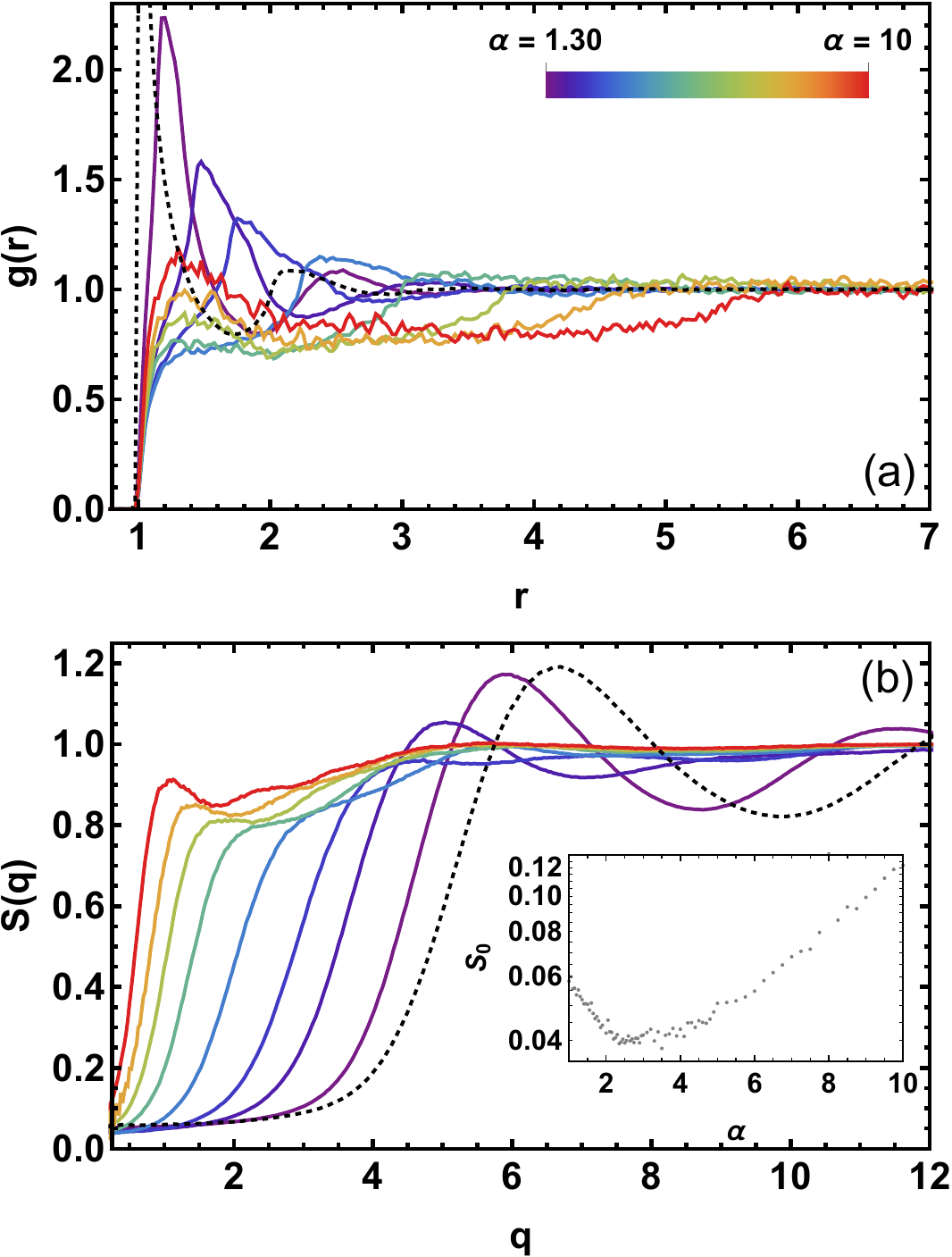}
\caption{Pair correlation functions $g(r)$ [panel (a)] and structure factors $S(q)$ [panel (b)] for the same systems highlighted in Figure \ref{fig:phivst} { \cite{footSofQvec}}.  Dotted curves show results for disks.  The inset to panel (b) shows estimated $S_0 = \lim_{q \to 0} S(q)$, here defined as $S(0.25)$.}
\label{fig:gands}
\end{figure}

Next we turn to detailed analyses of these packings' positional and orientational order.
Figure \ref{fig:gands} shows the pair correlation functions $g(r)$ and structure factors $S(q)$ for the same eight aspect ratios highlighted in Fig.\ \ref{fig:phivst}.
Three major trends are apparent.

First, all results for $\alpha \leq 2.2$ are consistent with previous studies \cite{ciesla16,haiduk18}.
The $g(r)$ exhibit a primary nearest-neighbor peak at $r_{\rm nn} \simeq (\alpha + 1)/2$ and a much smaller second-nearest-neighbor peak at $r_{\rm sn} = 2 +  \mathcal{O}(\alpha)$. 
Both peaks rapidly broaden as $\alpha$ increases.
The primary peak's position indicates that it corresponds to tip-side contacts.
A similar peak is present in RSA packings of low-$\alpha$ spherocylinders; this feature is likely universal for packings of convex anisotropic objects \cite{ricci94,ciesla16}.
The peak at $r_{\rm sn}$ vanishes for $\alpha \gtrsim 1.9$; ellipses with $\alpha \gtrsim 1.9$'s failure to form  an approximately-isotropic second coordination shell during RSA helps explain why their $\phi_s$ is maximized at this $\alpha$.

Second, { for all $4 \lesssim \alpha \lesssim 8$,} $g(r) < 1$ for all $r < (\alpha + 1)/2$.
This ``correlation hole'' \cite{degennes79} can be understood in terms of two-body impenetrability constraints.
Suppose $\Delta \theta$ is the difference in the orientation angles of two ellipses.
If the ellipses are in contact, the maximal distance $r_{\rm max}$ between their centers ranges from $1$ for $\Delta\theta = 0$, which corresponds to side-to-side contacts of perfectly aligned ellipses, to $(\alpha + 1)/2$ for $\Delta\theta = 90^\circ$, which corresponds to tip-to-side contacts of perpendicularly oriented ellipses.
Therefore, for any pair of ellipses (contacting or not) with $r < (\alpha + 1)/2$, the allowed $\Delta \theta$ are restricted, increasingly so as $r$ decreases towards $1$.
This tends to reduce the number of ellipse pairs with center-to-center distances $r < (\alpha + 1)/2$ relative to the number that would be present in an ideal gas.
The correlation hole emerges when this effect overwhelms the natural tendency of particles in RSA packings to form coordination shells \cite{swendsen81}. 
{ It disappears when a side-to-side contact peak at $r \simeq r_{\rm ss}$ [defined by $g(r_{\rm ss}) > 1$] emerges as $\alpha$ increases beyond $8$.}

Third, examining the $S(q)$ curves reveals how density fluctuations evolve with increasing $\alpha$.
Low-$\alpha$ systems' $S(q)$ are qualitatively similar to disks' $S(q)$.
They have prominent peaks at $q \simeq 2\pi/r_{\rm nn}$ and $q \simeq 4\pi/r_{\rm nn}$, which respectively correspond to the first- and second-nearest neighbor peaks shown in Fig.\ \ref{fig:gands}(a).
Two features which are not readily apparent from Fig.\ \ref{fig:gands}(a) are:\ (i) intermediate-wavelength ($1 \lesssim q \lesssim 4$) density fluctuations increase monotonically with $\alpha$, and (ii) longer-wavelength density fluctuations increase rapidly -- as shown in the inset, roughly exponentially -- for $\alpha \gtrsim 5$.
We will show below that the second trend corresponds to the emergence of lamellae that grow larger with increasing $\alpha$.

\begin{figure*}[htbp]
\includegraphics[width=7in]{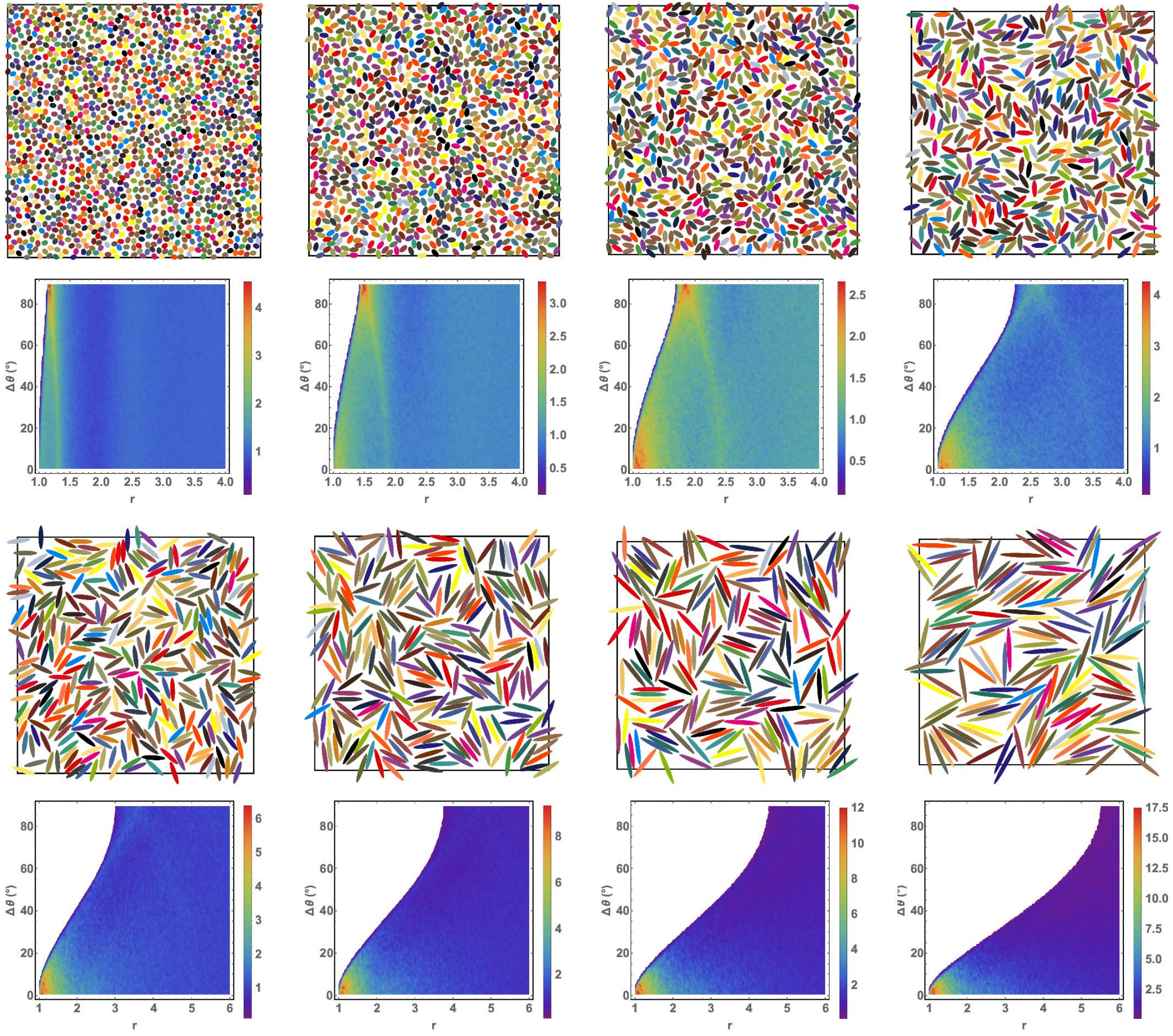}
\caption{Representative snapshots and $g(r,\Delta\theta)$ for $\alpha = 1.3$, $1.85$, $2.4$, and $3.5$ (top rows) and $\alpha = 5$, $6.5$, $8$, and $10$ (bottom rows) from left to right.  The snapshots show only ellipses whose \textit{centers} lie within the central $50\times 50$ regions of the overall $L\times L \equiv 1000\times 1000$ domains.}
\label{fig:snaps}
\end{figure*}

The abovementioned trends can be better understood by visualizing saturated RSA packings and their positional-orientational order.
Figure \ref{fig:snaps} shows snapshots of typical packings for the eight $\alpha$ highlighted above, along with results for the pair correlation function $g(r,\Delta\theta)$, which is the ratio of the number of ellipse pairs with center-to-center distance $r$ and orientation-angle difference $\Delta\theta$ to the number that would be present in an ideal gas of these particles.
In other words $g(r,\Delta\theta)$ is just the generalization of the standard pair correlation function $g(r)$ to include orientation-angle differences { \cite{footSofQvec}}.

Results for $\alpha \lesssim 2$ are consistent with previous studies \cite{sherwood90,viot92,ciesla16,haiduk18}, but shed new light on low-aspect-ratio RSA ellipse packings' structure because these systems'  $g(r,\Delta\theta)$ has not been previously reported.
The snapshots show the expected excess of tip-side contacts, and the $g(r,\Delta\theta)$ plots illustrate the degree to which this favored ordering diminishes { as} $r$ and $\Delta\theta$ vary away from the ideal tip-side configuration [$r_{\rm its} = (\alpha+1)/2,\ \Delta\theta = 90^\circ$].
The primary peaks in $g(r,\Delta\theta)$ occur at distances $r_{\rm ts} = r_{\rm its} + \mathcal{O}(\alpha)$ that increase with $\alpha$, but remain near $\Delta\theta = 90^\circ$.
Secondary peaks at distances $r_{\rm 2}(\alpha, \Delta\theta)$ that increase with increasing $\alpha$ and decreasing $\Delta\theta$, roughly mirroring the small-$r$ excluded-area cutoffs $r_{\rm min}(\alpha,\Delta\theta)$ below which $g(r,\Delta\theta) = 0$, are also evident.
These reflect the fact that there are at least two distinct favored near-neighbor configurations for any given $\Delta\theta$.

The snapshot and $g(r,\Delta\theta)$-plot for $\alpha = 3.5$ illustrate how the structures of saturated ellipse packings with $\alpha = 3-4$ differ from their lower-aspect-ratio counterparts.
Local nematic precursors comparable to those observed in both previous Monte Carlo simulations of hard ellipses and spherocylinders \cite{ricci94,xu13b} and experiments on quasi-2D colloidal-ellipsoid suspensions  \cite{zheng11,zheng14,zheng21,tan21} are readily visible.
Such precursor domains are believed to be largely responsible for these systems unusual liquid-state dynamics, and in particular for their ability to exhibit separate rotational and translational glass transitions \cite{wang18b,zheng11,zheng14,zheng21,roller21}.

The bottom rows of Fig.\ \ref{fig:snaps} show that for $\alpha \gtrsim 5$, saturated RSA ellipse packings possess a common structure that evolves quantitatively but not qualitatively with increasing $\alpha$.
These packings' $g(r,\Delta\theta)$ are simpler, with little structure aside from their side-to-side-contact peaks near $(r_{\rm ss},0)$.
The snapshots indicate that packings are largely composed of randomly-oriented, single-layer lamellae.
In contrast with the longer-range nematic order that develops in { sufficiently} dense systems of ellipses and other anisotropic particles in thermodynamic equilibrium \cite{ricci94,xu13b,bautista14b}, these lamellae appear have a kinetic origin; see below for a potential explanation. 
The combination of random lamellar orientation and increasing lamellar width (i.e.\ increasing particle length) with increasing $\alpha$ directly explains these packings' rapidly increasing long-wavelength density fluctuations [Fig.\ \ref{fig:gands}(b)].
Local particle density is high inside lamellae and low where lamellae of different orientations intersect (or, more precisely, where they would intersect if they were able to grow further).
Since (i) the size { of the lamellae} and (ii) the density disparity between the lamellae and their intersection regions both increase rapidly with $\alpha$, so does $S_0$.
The same effect is seen in RSA of zero-thickness line segments \cite{ziff90}.

One natural order parameter for these packings is $g_{\rm max} = \max[g(r,\Delta\theta)]$, the maximal excess positional-orientational correlation, here defined to occur for pairs with  center-to-center distance $r_{\rm max}$ and orientation-angle difference $\Delta\theta_{\rm max}$.
Comparing the snapshots and $g(r,\Delta\theta)$ shown in the top rows of Fig.\ \ref{fig:snaps} highlights our central result:\ the existence of a previously-unreported transition in RSA ellipse packings' structure.
 $g_{\rm max}$ decreases rapidly with increasing $\alpha$ over the range $1 < \alpha \leq \alpha_{\rm TS} \simeq 2.4$ as the peaks in $g(r)$ and $g(r,\Delta\theta)$ [respectively at $r_{\rm nn}$ and $r_{\rm ts}$] broaden and shift outward.
As $\alpha$ increases beyond $\alpha_{\rm TS}$, the positions of the peaks in $g(r,\Delta\theta)$ change discontinuously.
Specifically, $r_{\rm max}$ decreases from $r_{\rm ts} \simeq (\alpha+1)/2$ to $r_{\rm ss} \simeq 1.1$, and $\Delta\theta_{\rm max}$ decreases from just below $90^\circ$ to just above $0^\circ$.
Although the heights of the peaks in  $g(r,\Delta\theta)$ at $r_{\rm ts}$ and $r_{\rm ss}$ both evolve continuously with increasing $\alpha$, this discontinuous change in ($r_{\rm max}$, $\Delta\theta_{\rm max}$) can be regarded as a structural transition from tip-side- to side-side-dominated contact { \cite{footTStrans}}.

\begin{figure}[htbp]
\includegraphics[width=3in]{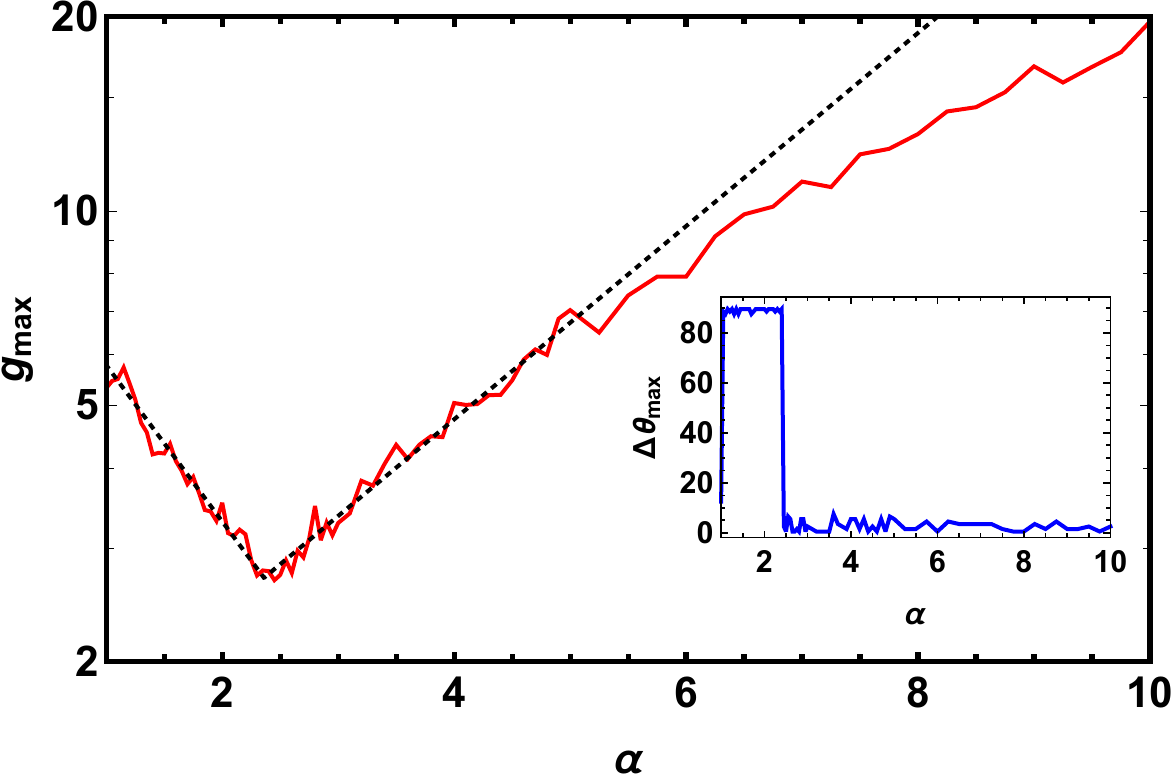}
\caption{ Exponential increase in maximal positional-orientational correlations with increasing $|\alpha-\alpha_{\rm TS}|$.  Dotted lines show Eq.\ \ref{eq:fitg}.  The inset shows the orientation-angle differences $\Delta\theta_{\rm max}$ at which the maxima occur, highlighting the transition from tip-side- to side-side-dominated contact.}
\label{fig:gmax}
\end{figure}

For $\alpha > \alpha_{\rm TS}$, $g_{\rm max}$ increases rapidly with $\alpha$ as side-to-side nearest-neighbor contacts with $r \simeq r_{\rm ss}$ become increasingly prevalent.
We find that in fact $g_{\rm max}$ increases exponentially with $|\alpha - \alpha_{\rm TS}|$.
As shown in Figure \ref{fig:gmax},  $g_{\rm max}(\alpha)$ is well fit by
\begin{equation}
g_{\rm max} \simeq \Bigg{ \{ } \begin{array}{lcr}
g_0 \exp\left( \displaystyle\frac{\alpha_{\rm TS}^* - \alpha}{\alpha^*_{\rm low} } \right) & , & \alpha < \alpha_{\rm TS} \\
\\
g_0 \exp\left( \displaystyle\frac{\alpha_{\rm TS}^* - \alpha}{ \alpha^*_{\rm high} }  \right) & , & \alpha > \alpha_{\rm TS} 
\end{array},
\label{eq:fitg}
\end{equation}
for all $\alpha \lesssim 5$, with $g_0 \simeq 2.70$, $\alpha_{\rm TS}^* \simeq 2.36$, $\alpha^*_{\rm low} \simeq 1.79$ and $\alpha^*_{\rm high} \simeq 2.90$.
Systems with $\alpha = \alpha_{\rm TS}^*$ can be considered maximally { locally} disordered.
They are frustrated in the sense that tip-side- and side-side-contact-dominated ordering are mutually incompatible { yet equally likely.}

While the values of the parameters $g_0$, $\alpha_{\rm TS}^*$, $\alpha^*_{\rm low}$ and $\alpha^*_{\rm high}$ depend on how the binning used to calculate $g(r,\Delta\theta)$ is performed, the functional form of Eq.\ \ref{eq:fitg} is robust.
Rigorously explaining this result is challenging.
In particular, two-body geometry alone does not explain it, because although the range of $\Delta\theta$ [$0 \leq \Delta\theta \leq \Delta\theta^*(r)$] over which a new ellipse can be inserted with a center-to-center distance $r \lesssim r_{\rm ss}$ from its nearest neighbor decreases with increasing $\alpha$, this decrease would only produce $g_{\rm max} \sim 90^\circ/\Delta\theta^*(r) \sim \alpha$.
We suspect that, at least for $\alpha > \alpha_{\rm TS}$, the exponential increase of $g_{\rm max}$ arises from a combination of geometry and kinetics.
The locally densest configurations of ellipses with $r \simeq r_{\rm ss}$ are single-layer lamellae.
As RSA proceeds, the gaps into which new particles may be inserted increasingly correspond to configurations that increase the length of such lamellae, i.e.\ grow the lamellae at their ends.
Thus $g_{\rm max}$ increases much faster than linearly with $\alpha$, falling below its initial exponential trend only when lamellar growth becomes limited by their intersection with other, differently oriented lamellae, i.e.\ at $\alpha \simeq 5$.
{ For $\alpha > 5$ the growth of $g_{\rm max}$ with $\alpha$ appears to be linear.}

\section{Discussion and Conclusions}

Hard ellipse liquids' dynamics exhibit multiple nontrivial features arising from the asymmetry of their constituent particles \cite{zheng11,zheng14,mishra13,zheng21,shen12,xu15,wang18b,shen12,davatolhagh12}.
Thus they provide a minimal, single-parameter model for understanding aspects of the  dynamics of small-molecule glassforming liquids that are tightly coupled to particle asymmetry, such as translational-rotational decoupling \cite{chong09}.
They also provide an easily-visualized system that can be used as a basis for understanding their more complicated three-dimensional-ellipsoidal counterparts \cite{mishra14,mishra15,roller21,roller20,pal20,pal21}.

In this paper, we characterized the structure of asymptotically saturated RSA ellipse packings, over a far wider range of aspect ratios and in substantially more detail than had been previously reported.
We obtained a simple analytic expression for the saturation density $\phi_s(\alpha)$ [Eq.\ \ref{eq:analyticphis}] that captures results for all $\alpha \leq 10$ to within less than $0.1\%$ and has the correct $1/\alpha$ scaling \cite{philipse96} in the large-$\alpha$ limit.
The crossover to this asymptotic scaling was slower than expected; while it must eventually occur since $\phi_s$ is necessarily below the jamming density $\phi_J(\alpha)$ \cite{donev04,donev07,delaney05,schreck09}, we showed that it does so only at some $\alpha \gg 10$.

We also characterized how positional-orientational correlations in these packings evolve with increasing $\alpha$.
Our main result for relatively low aspect rations ($\alpha \leq 5$) was finding a previously-unreported structural transition from tip/side- to side/side-contact-dominated structure at $\alpha = \alpha_{\rm TS} \simeq 2.4$.
At this aspect ratio, the peak value $g_{\rm max}$ of packings' positional-orientational pair correlation function $g(r,\Delta\theta)$ is minimal, and packings can be considered maximally frustrated owing to competition of these two, mutually incompatible types of structural order.
$g_{\rm max}$ increases exponentially with increasing $|\alpha - \alpha_{\rm TS}|$, presumably owing primarily to kinetic effects.

{ $\alpha \simeq 2.4$ is also the aspect ratio above which ellipses possess an equilibrium nematic liquid-crystalline phase \cite{bautista14b}, suggesting that the abovementioned structural transition is intimately connected to the equilibrium isotropic-nematic transition.
Explaining such a connection -- if indeed one exists -- will be nontrivial because saturated RSA packings of anistropic 2D particles often have local structure that is substantially different than that of their equilibrium counterparts at the same $\phi$ \cite{ricci94}.
One potential strategy for obtaining such an explanation is extending replica-trick-based techniques, which were very recently employed to successfully predict RSA sphere packings'  structure \cite{jadrich22}, to anisotropic particles.}

RSA ellipse packings with $\alpha > 5$ had not been previously investigated.
We found that these packings possess a common structure that varies quantitatively but not qualitatively with increasing $\alpha$:\ they are primarily composed of randomly oriented single-layer lamellae.
$g_{\rm max}$ falls below the abovementioned exponential trend, presumably because  lamellar growth is limited by ``collisions'' with other, differently oriented lamellae.
The same collisions drive a rapid increase in long-wavelength density fluctuations as the size of and density disparity between lamellae and their intersection regions increase with increasing $\alpha$.

In conclusion, the detailed characterization of saturated RSA ellipse packings performed in this work, in addition to being of interest in its own right, lays the groundwork for followup studies of the structure and dynamics of thermally-equilibrated, glassy and jammed ellipses.
For example, 
it would be interesting to check whether a comparable structural transition and comparable behavior of $g_{\rm max}$ are present in jammed ellipse packings, and if so, at which aspect ratio it occurs.
Our results suggest that followup studies examining  $g(r,{\Delta}\theta)$ and/or systems with $\alpha > 5$ may prove especially fruitful.

This material is based upon work supported by the National Science Foundation under Grant No.\ DMR-2026271.


\begin{thebibliography}{47}%
\makeatletter
\providecommand \@ifxundefined [1]{%
 \@ifx{#1\undefined}
}%
\providecommand \@ifnum [1]{%
 \ifnum #1\expandafter \@firstoftwo
 \else \expandafter \@secondoftwo
 \fi
}%
\providecommand \@ifx [1]{%
 \ifx #1\expandafter \@firstoftwo
 \else \expandafter \@secondoftwo
 \fi
}%
\providecommand \natexlab [1]{#1}%
\providecommand \enquote  [1]{``#1''}%
\providecommand \bibnamefont  [1]{#1}%
\providecommand \bibfnamefont [1]{#1}%
\providecommand \citenamefont [1]{#1}%
\providecommand \href@noop [0]{\@secondoftwo}%
\providecommand \href [0]{\begingroup \@sanitize@url \@href}%
\providecommand \@href[1]{\@@startlink{#1}\@@href}%
\providecommand \@@href[1]{\endgroup#1\@@endlink}%
\providecommand \@sanitize@url [0]{\catcode `\\12\catcode `\$12\catcode
  `\&12\catcode `\#12\catcode `\^12\catcode `\_12\catcode `\%12\relax}%
\providecommand \@@startlink[1]{}%
\providecommand \@@endlink[0]{}%
\providecommand \url  [0]{\begingroup\@sanitize@url \@url }%
\providecommand \@url [1]{\endgroup\@href {#1}{\urlprefix }}%
\providecommand \urlprefix  [0]{URL }%
\providecommand \Eprint [0]{\href }%
\providecommand \doibase [0]{http://dx.doi.org/}%
\providecommand \selectlanguage [0]{\@gobble}%
\providecommand \bibinfo  [0]{\@secondoftwo}%
\providecommand \bibfield  [0]{\@secondoftwo}%
\providecommand \translation [1]{[#1]}%
\providecommand \BibitemOpen [0]{}%
\providecommand \bibitemStop [0]{}%
\providecommand \bibitemNoStop [0]{.\EOS\space}%
\providecommand \EOS [0]{\spacefactor3000\relax}%
\providecommand \BibitemShut  [1]{\csname bibitem#1\endcsname}%
\let\auto@bib@innerbib\@empty
\bibitem [{\citenamefont {Dugyala}\ and\ \citenamefont {amd
  M.~G.~Basavaraj}(2013)}]{dugyala13}%
  \BibitemOpen
  \bibfield  {author} {\bibinfo {author} {\bibfnamefont {V.~R.}\ \bibnamefont
  {Dugyala}}\ and\ \bibinfo {author} {\bibfnamefont {S.~V.~Daware}\
  \bibnamefont {amd M.~G.~Basavaraj}},\ }\bibfield  {title} {\enquote {\bibinfo
  {title} {Shape anisotropic colloids: synthesis, packing behavior, evaporation
  driven assembly, and their application in emulsion stabilization},}\
  }\href@noop {} {\bibfield  {journal} {\bibinfo  {journal} {Soft Matt.}\
  }\textbf {\bibinfo {volume} {9}},\ \bibinfo {pages} {6711} (\bibinfo {year}
  {2013})}\BibitemShut {NoStop}%
\bibitem [{\citenamefont {Weeks}\ \emph {et~al.}(2000)\citenamefont {Weeks},
  \citenamefont {Crocker}, \citenamefont {Levitt}, \citenamefont {Schofield},\
  and\ \citenamefont {Weitz}}]{weeks00}%
  \BibitemOpen
  \bibfield  {author} {\bibinfo {author} {\bibfnamefont {E.~R.}\ \bibnamefont
  {Weeks}}, \bibinfo {author} {\bibfnamefont {J.~C.}\ \bibnamefont {Crocker}},
  \bibinfo {author} {\bibfnamefont {A.~C.}\ \bibnamefont {Levitt}}, \bibinfo
  {author} {\bibfnamefont {A.}~\bibnamefont {Schofield}}, \ and\ \bibinfo
  {author} {\bibfnamefont {D.~A.}\ \bibnamefont {Weitz}},\ }\bibfield  {title}
  {\enquote {\bibinfo {title} {Three-dimensional direct imaging of structural
  relaxation near the colloidal glass transition},}\ }\href@noop {} {\bibfield
  {journal} {\bibinfo  {journal} {Science}\ }\textbf {\bibinfo {volume}
  {287}},\ \bibinfo {pages} {627} (\bibinfo {year} {2000})}\BibitemShut
  {NoStop}%
\bibitem [{\citenamefont {Gasser}\ \emph {et~al.}(2001)\citenamefont {Gasser},
  \citenamefont {Weeks}, \citenamefont {Schofield}, \citenamefont {Pusey},\
  and\ \citenamefont {Weitz}}]{gasser01}%
  \BibitemOpen
  \bibfield  {author} {\bibinfo {author} {\bibfnamefont {U.}~\bibnamefont
  {Gasser}}, \bibinfo {author} {\bibfnamefont {E.~R.}\ \bibnamefont {Weeks}},
  \bibinfo {author} {\bibfnamefont {A.}~\bibnamefont {Schofield}}, \bibinfo
  {author} {\bibfnamefont {P.~N.}\ \bibnamefont {Pusey}}, \ and\ \bibinfo
  {author} {\bibfnamefont {D.~A.}\ \bibnamefont {Weitz}},\ }\bibfield  {title}
  {\enquote {\bibinfo {title} {Real-space imaging of nucleation and growth in
  colloidal crystallization},}\ }\href@noop {} {\bibfield  {journal} {\bibinfo
  {journal} {Science}\ }\textbf {\bibinfo {volume} {292}},\ \bibinfo {pages}
  {258} (\bibinfo {year} {2001})}\BibitemShut {NoStop}%
\bibitem [{\citenamefont {Roller}\ \emph {et~al.}(2021)\citenamefont {Roller},
  \citenamefont {Laganapan}, \citenamefont {Meijer}, \citenamefont {Fuchs},\
  and\ \citenamefont {Zumbusch}}]{roller21}%
  \BibitemOpen
  \bibfield  {author} {\bibinfo {author} {\bibfnamefont {J.}~\bibnamefont
  {Roller}}, \bibinfo {author} {\bibfnamefont {A.}~\bibnamefont {Laganapan}},
  \bibinfo {author} {\bibfnamefont {J.-M.}\ \bibnamefont {Meijer}}, \bibinfo
  {author} {\bibfnamefont {M.}~\bibnamefont {Fuchs}}, \ and\ \bibinfo {author}
  {\bibfnamefont {A.}~\bibnamefont {Zumbusch}},\ }\bibfield  {title} {\enquote
  {\bibinfo {title} {Observation of liquid glass in suspensions of ellipsoidal
  colloids},}\ }\href@noop {} {\bibfield  {journal} {\bibinfo  {journal} {Proc.
  Nat. Acad. Sci.}\ }\textbf {\bibinfo {volume} {118}},\ \bibinfo {pages}
  {2018072118} (\bibinfo {year} {2021})}\BibitemShut {NoStop}%
\bibitem [{\citenamefont {Letz}\ \emph {et~al.}(2000)\citenamefont {Letz},
  \citenamefont {Schilling},\ and\ \citenamefont {Latz}}]{letz00}%
  \BibitemOpen
  \bibfield  {author} {\bibinfo {author} {\bibfnamefont {M.}~\bibnamefont
  {Letz}}, \bibinfo {author} {\bibfnamefont {R.}~\bibnamefont {Schilling}}, \
  and\ \bibinfo {author} {\bibfnamefont {A.}~\bibnamefont {Latz}},\ }\bibfield
  {title} {\enquote {\bibinfo {title} {Ideal glass transitions for hard
  ellipsoids},}\ }\href@noop {} {\bibfield  {journal} {\bibinfo  {journal}
  {Phys. Rev. E}\ }\textbf {\bibinfo {volume} {62}},\ \bibinfo {pages} {5173}
  (\bibinfo {year} {2000})}\BibitemShut {NoStop}%
\bibitem [{\citenamefont {Zheng}\ \emph {et~al.}(2011)\citenamefont {Zheng},
  \citenamefont {Wang},\ and\ \citenamefont {Han}}]{zheng11}%
  \BibitemOpen
  \bibfield  {author} {\bibinfo {author} {\bibfnamefont {Z.}~\bibnamefont
  {Zheng}}, \bibinfo {author} {\bibfnamefont {F.}~\bibnamefont {Wang}}, \ and\
  \bibinfo {author} {\bibfnamefont {Y.}~\bibnamefont {Han}},\ }\bibfield
  {title} {\enquote {\bibinfo {title} {Glass transitions in
  quasi-two-dimensional suspensions of colloidal ellipsoids},}\ }\href@noop {}
  {\bibfield  {journal} {\bibinfo  {journal} {Phys. Rev. Lett.}\ }\textbf
  {\bibinfo {volume} {107}},\ \bibinfo {pages} {065702} (\bibinfo {year}
  {2011})}\BibitemShut {NoStop}%
\bibitem [{\citenamefont {Zheng}\ \emph {et~al.}(2014)\citenamefont {Zheng},
  \citenamefont {Ni}, \citenamefont {Wang}, \citenamefont {Dijkstra},
  \citenamefont {Wang},\ and\ \citenamefont {Han}}]{zheng14}%
  \BibitemOpen
  \bibfield  {author} {\bibinfo {author} {\bibfnamefont {Z.}~\bibnamefont
  {Zheng}}, \bibinfo {author} {\bibfnamefont {R.}~\bibnamefont {Ni}}, \bibinfo
  {author} {\bibfnamefont {F.}~\bibnamefont {Wang}}, \bibinfo {author}
  {\bibfnamefont {M.}~\bibnamefont {Dijkstra}}, \bibinfo {author}
  {\bibfnamefont {Y.}~\bibnamefont {Wang}}, \ and\ \bibinfo {author}
  {\bibfnamefont {Y.}~\bibnamefont {Han}},\ }\bibfield  {title} {\enquote
  {\bibinfo {title} {Structural signatures of dynamic heterogeneities in
  monolayers of colloidal ellipsoids},}\ }\href@noop {} {\bibfield  {journal}
  {\bibinfo  {journal} {Nat. Comm.}\ }\textbf {\bibinfo {volume} {5}},\
  \bibinfo {pages} {3829} (\bibinfo {year} {2014})}\BibitemShut {NoStop}%
\bibitem [{\citenamefont {Mishra}\ \emph {et~al.}(2013)\citenamefont {Mishra},
  \citenamefont {Rangarajan},\ and\ \citenamefont {Ganapathy}}]{mishra13}%
  \BibitemOpen
  \bibfield  {author} {\bibinfo {author} {\bibfnamefont {C.~K.}\ \bibnamefont
  {Mishra}}, \bibinfo {author} {\bibfnamefont {A.}~\bibnamefont {Rangarajan}},
  \ and\ \bibinfo {author} {\bibfnamefont {R.}~\bibnamefont {Ganapathy}},\
  }\bibfield  {title} {\enquote {\bibinfo {title} {Two-step glass transition
  induced by attractive interactions in quasi-two-dimensional suspensions of
  ellipsoidal particles},}\ }\href@noop {} {\bibfield  {journal} {\bibinfo
  {journal} {Phys. Rev. Lett.}\ }\textbf {\bibinfo {volume} {110}},\ \bibinfo
  {pages} {188301} (\bibinfo {year} {2013})}\BibitemShut {NoStop}%
\bibitem [{\citenamefont {Sastry}\ \emph {et~al.}(1998)\citenamefont {Sastry},
  \citenamefont {Debenedetti}, \citenamefont {Torquato},\ and\ \citenamefont
  {Stillinger}}]{sastry98}%
  \BibitemOpen
  \bibfield  {author} {\bibinfo {author} {\bibfnamefont {S.}~\bibnamefont
  {Sastry}}, \bibinfo {author} {\bibfnamefont {P.~G.}\ \bibnamefont
  {Debenedetti}}, \bibinfo {author} {\bibfnamefont {S.}~\bibnamefont
  {Torquato}}, \ and\ \bibinfo {author} {\bibfnamefont {F.~H.}\ \bibnamefont
  {Stillinger}},\ }\bibfield  {title} {\enquote {\bibinfo {title} {Signatures
  of distinct dynamical regimes in the energy landscape of a glass-forming
  liquid},}\ }\href@noop {} {\bibfield  {journal} {\bibinfo  {journal}
  {Nature}\ }\textbf {\bibinfo {volume} {393}},\ \bibinfo {pages} {554}
  (\bibinfo {year} {1998})}\BibitemShut {NoStop}%
\bibitem [{\citenamefont {Chaudhuri}\ \emph {et~al.}(2010)\citenamefont
  {Chaudhuri}, \citenamefont {Berthier},\ and\ \citenamefont
  {Sastry}}]{chaudhuri10}%
  \BibitemOpen
  \bibfield  {author} {\bibinfo {author} {\bibfnamefont {P.}~\bibnamefont
  {Chaudhuri}}, \bibinfo {author} {\bibfnamefont {L.}~\bibnamefont {Berthier}},
  \ and\ \bibinfo {author} {\bibfnamefont {S.}~\bibnamefont {Sastry}},\
  }\bibfield  {title} {\enquote {\bibinfo {title} {Jamming transitions in
  amorphous packings of frictionless spheres occur over a continuous range of
  volume fractions},}\ }\href@noop {} {\bibfield  {journal} {\bibinfo
  {journal} {Phys. Rev. Lett.}\ }\textbf {\bibinfo {volume} {104}},\ \bibinfo
  {pages} {165701} (\bibinfo {year} {2010})}\BibitemShut {NoStop}%
\bibitem [{\citenamefont {Pfleiderer}\ \emph {et~al.}(2008)\citenamefont
  {Pfleiderer}, \citenamefont {Milinkovic},\ and\ \citenamefont
  {Schilling}}]{pfleiderer08}%
  \BibitemOpen
  \bibfield  {author} {\bibinfo {author} {\bibfnamefont {P.}~\bibnamefont
  {Pfleiderer}}, \bibinfo {author} {\bibfnamefont {K.}~\bibnamefont
  {Milinkovic}}, \ and\ \bibinfo {author} {\bibfnamefont {T.}~\bibnamefont
  {Schilling}},\ }\bibfield  {title} {\enquote {\bibinfo {title} {Glassy
  dynamics in monodisperse hard ellipsoids},}\ }\href@noop {} {\bibfield
  {journal} {\bibinfo  {journal} {Europhys. Lett.}\ }\textbf {\bibinfo {volume}
  {84}},\ \bibinfo {pages} {16003} (\bibinfo {year} {2008})}\BibitemShut
  {NoStop}%
\bibitem [{\citenamefont {Shen}\ \emph {et~al.}(2012)\citenamefont {Shen},
  \citenamefont {Schreck}, \citenamefont {Chakraborty}, \citenamefont {Freed},\
  and\ \citenamefont {O'Hern}}]{shen12}%
  \BibitemOpen
  \bibfield  {author} {\bibinfo {author} {\bibfnamefont {T.}~\bibnamefont
  {Shen}}, \bibinfo {author} {\bibfnamefont {C.}~\bibnamefont {Schreck}},
  \bibinfo {author} {\bibfnamefont {B.}~\bibnamefont {Chakraborty}}, \bibinfo
  {author} {\bibfnamefont {D.~E.}\ \bibnamefont {Freed}}, \ and\ \bibinfo
  {author} {\bibfnamefont {C.~S.}\ \bibnamefont {O'Hern}},\ }\bibfield  {title}
  {\enquote {\bibinfo {title} {Structural relaxation in dense liquids composed
  of anisotropic particles},}\ }\href@noop {} {\bibfield  {journal} {\bibinfo
  {journal} {Phys. Rev. E}\ }\textbf {\bibinfo {volume} {86}},\ \bibinfo
  {pages} {041303} (\bibinfo {year} {2012})}\BibitemShut {NoStop}%
\bibitem [{\citenamefont {Davatolhagh}\ and\ \citenamefont
  {Foroozan}(2012)}]{davatolhagh12}%
  \BibitemOpen
  \bibfield  {author} {\bibinfo {author} {\bibfnamefont {S.}~\bibnamefont
  {Davatolhagh}}\ and\ \bibinfo {author} {\bibfnamefont {S.}~\bibnamefont
  {Foroozan}},\ }\bibfield  {title} {\enquote {\bibinfo {title} {Structural
  origin of enhanced translational diffusion in two-dimensional hard-ellipse
  fluids},}\ }\href@noop {} {\bibfield  {journal} {\bibinfo  {journal} {Phys.
  Rev. E}\ }\textbf {\bibinfo {volume} {85}},\ \bibinfo {pages} {061707}
  (\bibinfo {year} {2012})}\BibitemShut {NoStop}%
\bibitem [{\citenamefont {Xu}\ \emph {et~al.}(2015)\citenamefont {Xu},
  \citenamefont {Sun},\ and\ \citenamefont {An}}]{xu15}%
  \BibitemOpen
  \bibfield  {author} {\bibinfo {author} {\bibfnamefont {W.-S.}\ \bibnamefont
  {Xu}}, \bibinfo {author} {\bibfnamefont {Z.-Y.}\ \bibnamefont {Sun}}, \ and\
  \bibinfo {author} {\bibfnamefont {L.-J.}\ \bibnamefont {An}},\ }\bibfield
  {title} {\enquote {\bibinfo {title} {Relaxation dynamics in a binary
  hard-ellipse liquid},}\ }\href@noop {} {\bibfield  {journal} {\bibinfo
  {journal} {Soft Matt.}\ }\textbf {\bibinfo {volume} {11}},\ \bibinfo {pages}
  {627} (\bibinfo {year} {2015})}\BibitemShut {NoStop}%
\bibitem [{\citenamefont {Philipse}(1996)}]{philipse96}%
  \BibitemOpen
  \bibfield  {author} {\bibinfo {author} {\bibfnamefont {A.~P.}\ \bibnamefont
  {Philipse}},\ }\bibfield  {title} {\enquote {\bibinfo {title} {The random
  contact equation and its implications for (colloidal) rods in packings,
  suspensions, and anisotropic powders},}\ }\href@noop {} {\bibfield  {journal}
  {\bibinfo  {journal} {Langmuir}\ }\textbf {\bibinfo {volume} {12}},\ \bibinfo
  {pages} {1127} (\bibinfo {year} {1996})}\BibitemShut {NoStop}%
\bibitem [{\citenamefont {Desmond}\ and\ \citenamefont
  {Franklin}(2006)}]{desmond06}%
  \BibitemOpen
  \bibfield  {author} {\bibinfo {author} {\bibfnamefont {K.}~\bibnamefont
  {Desmond}}\ and\ \bibinfo {author} {\bibfnamefont {S.~V.}\ \bibnamefont
  {Franklin}},\ }\bibfield  {title} {\enquote {\bibinfo {title} {Jamming of
  three-dimensional prolate granular materials},}\ }\href@noop {} {\bibfield
  {journal} {\bibinfo  {journal} {Phys. Rev. E}\ }\textbf {\bibinfo {volume}
  {73}},\ \bibinfo {pages} {031306} (\bibinfo {year} {2006})}\BibitemShut
  {NoStop}%
\bibitem [{\citenamefont {Donev}\ \emph {et~al.}(2004)\citenamefont {Donev},
  \citenamefont {Cisse}, \citenamefont {Sachs}, \citenamefont {Variano},
  \citenamefont {Stillinger}, \citenamefont {Connelly}, \citenamefont
  {Torquato},\ and\ \citenamefont {Chaikin}}]{donev04}%
  \BibitemOpen
  \bibfield  {author} {\bibinfo {author} {\bibfnamefont {A.}~\bibnamefont
  {Donev}}, \bibinfo {author} {\bibfnamefont {I.}~\bibnamefont {Cisse}},
  \bibinfo {author} {\bibfnamefont {D.}~\bibnamefont {Sachs}}, \bibinfo
  {author} {\bibfnamefont {E.~A.}\ \bibnamefont {Variano}}, \bibinfo {author}
  {\bibfnamefont {F.~H.}\ \bibnamefont {Stillinger}}, \bibinfo {author}
  {\bibfnamefont {R.}~\bibnamefont {Connelly}}, \bibinfo {author}
  {\bibfnamefont {S.}~\bibnamefont {Torquato}}, \ and\ \bibinfo {author}
  {\bibfnamefont {P.~M.}\ \bibnamefont {Chaikin}},\ }\bibfield  {title}
  {\enquote {\bibinfo {title} {Improving the density of jammed disordered
  packings using ellipsoids},}\ }\href@noop {} {\bibfield  {journal} {\bibinfo
  {journal} {Science}\ }\textbf {\bibinfo {volume} {303}},\ \bibinfo {pages}
  {990} (\bibinfo {year} {2004})}\BibitemShut {NoStop}%
\bibitem [{\citenamefont {Donev}\ \emph {et~al.}(2007)\citenamefont {Donev},
  \citenamefont {Connelly}, \citenamefont {Stillinger},\ and\ \citenamefont
  {Torquato}}]{donev07}%
  \BibitemOpen
  \bibfield  {author} {\bibinfo {author} {\bibfnamefont {A.}~\bibnamefont
  {Donev}}, \bibinfo {author} {\bibfnamefont {R.}~\bibnamefont {Connelly}},
  \bibinfo {author} {\bibfnamefont {F.~H.}\ \bibnamefont {Stillinger}}, \ and\
  \bibinfo {author} {\bibfnamefont {S.}~\bibnamefont {Torquato}},\ }\bibfield
  {title} {\enquote {\bibinfo {title} {Underconstrained jammed packings of
  nonspherical hard particles: Ellipses and ellipsoids},}\ }\href@noop {}
  {\bibfield  {journal} {\bibinfo  {journal} {Phys. Rev. E}\ }\textbf {\bibinfo
  {volume} {75}},\ \bibinfo {pages} {051304} (\bibinfo {year}
  {2007})}\BibitemShut {NoStop}%
\bibitem [{\citenamefont {Delaney}\ \emph {et~al.}(2006)\citenamefont
  {Delaney}, \citenamefont {Weaire}, \citenamefont {Hutzler},\ and\
  \citenamefont {Murphy}}]{delaney05}%
  \BibitemOpen
  \bibfield  {author} {\bibinfo {author} {\bibfnamefont {G.}~\bibnamefont
  {Delaney}}, \bibinfo {author} {\bibfnamefont {D.}~\bibnamefont {Weaire}},
  \bibinfo {author} {\bibfnamefont {S.}~\bibnamefont {Hutzler}}, \ and\
  \bibinfo {author} {\bibfnamefont {S.}~\bibnamefont {Murphy}},\ }\bibfield
  {title} {\enquote {\bibinfo {title} {Random packing of elliptical disks},}\
  }\href@noop {} {\bibfield  {journal} {\bibinfo  {journal} {Phil. Mag. Lett.}\
  }\textbf {\bibinfo {volume} {85}},\ \bibinfo {pages} {89} (\bibinfo {year}
  {2006})}\BibitemShut {NoStop}%
\bibitem [{\citenamefont {Schreck}\ \emph {et~al.}(2009)\citenamefont
  {Schreck}, \citenamefont {Xu},\ and\ \citenamefont {O'Hern}}]{schreck09}%
  \BibitemOpen
  \bibfield  {author} {\bibinfo {author} {\bibfnamefont {C.~F.}\ \bibnamefont
  {Schreck}}, \bibinfo {author} {\bibfnamefont {N.}~\bibnamefont {Xu}}, \ and\
  \bibinfo {author} {\bibfnamefont {C.~S.}\ \bibnamefont {O'Hern}},\ }\bibfield
   {title} {\enquote {\bibinfo {title} {A comparison of jamming behavior in
  systems composed of dimer- and ellipse-shaped particles},}\ }\href@noop {}
  {\bibfield  {journal} {\bibinfo  {journal} {Soft Matt.}\ }\textbf {\bibinfo
  {volume} {6}},\ \bibinfo {pages} {2960} (\bibinfo {year} {2009})}\BibitemShut
  {NoStop}%
\bibitem [{\citenamefont {Sherwood}(1990)}]{sherwood90}%
  \BibitemOpen
  \bibfield  {author} {\bibinfo {author} {\bibfnamefont {J.~D.}\ \bibnamefont
  {Sherwood}},\ }\bibfield  {title} {\enquote {\bibinfo {title} {Random
  sequential adsorption of lines and ellipses},}\ }\href@noop {} {\bibfield
  {journal} {\bibinfo  {journal} {J. Phys. A. Math. Gen.}\ }\textbf {\bibinfo
  {volume} {23}},\ \bibinfo {pages} {2827} (\bibinfo {year}
  {1990})}\BibitemShut {NoStop}%
\bibitem [{\citenamefont {Viot}\ \emph {et~al.}(1992)\citenamefont {Viot},
  \citenamefont {Tarjus}, \citenamefont {Ricci},\ and\ \citenamefont
  {Talbot}}]{viot92}%
  \BibitemOpen
  \bibfield  {author} {\bibinfo {author} {\bibfnamefont {P.}~\bibnamefont
  {Viot}}, \bibinfo {author} {\bibfnamefont {G.}~\bibnamefont {Tarjus}},
  \bibinfo {author} {\bibfnamefont {S.~M.}\ \bibnamefont {Ricci}}, \ and\
  \bibinfo {author} {\bibfnamefont {J.}~\bibnamefont {Talbot}},\ }\bibfield
  {title} {\enquote {\bibinfo {title} {Random sequential adsorption of
  anisotropic particles. {I}. {J}amming limit and asymptotic behavior},}\
  }\href@noop {} {\bibfield  {journal} {\bibinfo  {journal} {J. Chem. Phys.}\
  }\textbf {\bibinfo {volume} {97}},\ \bibinfo {pages} {5212} (\bibinfo {year}
  {1992})}\BibitemShut {NoStop}%
\bibitem [{\citenamefont {Cie{\'s}la}\ \emph {et~al.}(2016)\citenamefont
  {Cie{\'s}la}, \citenamefont {Pajak},\ and\ \citenamefont {Ziff}}]{ciesla16}%
  \BibitemOpen
  \bibfield  {author} {\bibinfo {author} {\bibfnamefont {M.}~\bibnamefont
  {Cie{\'s}la}}, \bibinfo {author} {\bibfnamefont {G.}~\bibnamefont {Pajak}}, \
  and\ \bibinfo {author} {\bibfnamefont {R.~M.}\ \bibnamefont {Ziff}},\
  }\bibfield  {title} {\enquote {\bibinfo {title} {In a search for a shape
  maximizing packing fraction for two-dimensional random sequential
  adsorption},}\ }\href@noop {} {\bibfield  {journal} {\bibinfo  {journal} {J.
  Chem. Phys.}\ }\textbf {\bibinfo {volume} {145}},\ \bibinfo {pages} {044708}
  (\bibinfo {year} {2016})}\BibitemShut {NoStop}%
\bibitem [{\citenamefont {Haiduk}\ \emph {et~al.}(2018)\citenamefont {Haiduk},
  \citenamefont {Kubala},\ and\ \citenamefont {Cie{\'s}la}}]{haiduk18}%
  \BibitemOpen
  \bibfield  {author} {\bibinfo {author} {\bibfnamefont {K.}~\bibnamefont
  {Haiduk}}, \bibinfo {author} {\bibfnamefont {P.}~\bibnamefont {Kubala}}, \
  and\ \bibinfo {author} {\bibfnamefont {M.}~\bibnamefont {Cie{\'s}la}},\
  }\bibfield  {title} {\enquote {\bibinfo {title} {Saturated packings of convex
  anisotropic objects under random sequential adsorption protocol},}\
  }\href@noop {} {\bibfield  {journal} {\bibinfo  {journal} {Phys. Rev. E}\
  }\textbf {\bibinfo {volume} {98}},\ \bibinfo {pages} {063309} (\bibinfo
  {year} {2018})}\BibitemShut {NoStop}%
\bibitem [{\citenamefont {Swendsen}(1981)}]{swendsen81}%
  \BibitemOpen
  \bibfield  {author} {\bibinfo {author} {\bibfnamefont {R.~H.}\ \bibnamefont
  {Swendsen}},\ }\bibfield  {title} {\enquote {\bibinfo {title} {Dynamics of
  random sequential adsorption},}\ }\href@noop {} {\bibfield  {journal}
  {\bibinfo  {journal} {Phys. Rev. A}\ }\textbf {\bibinfo {volume} {24}},\
  \bibinfo {pages} {504} (\bibinfo {year} {1981})}\BibitemShut {NoStop}%
\bibitem [{\citenamefont {Zheng}\ and\ \citenamefont
  {Palffy-Muhoray}(2007)}]{zheng07}%
  \BibitemOpen
  \bibfield  {author} {\bibinfo {author} {\bibfnamefont {X.}~\bibnamefont
  {Zheng}}\ and\ \bibinfo {author} {\bibfnamefont {P.}~\bibnamefont
  {Palffy-Muhoray}},\ }\bibfield  {title} {\enquote {\bibinfo {title} {Distance
  of closest approach of two arbitrary hard ellipses in two dimensions},}\
  }\href@noop {} {\bibfield  {journal} {\bibinfo  {journal} {Phys. Rev. E}\
  }\textbf {\bibinfo {volume} {75}},\ \bibinfo {pages} {061709} (\bibinfo
  {year} {2007})}\BibitemShut {NoStop}%
\bibitem [{\citenamefont {Vigil}\ and\ \citenamefont {Ziff}(1989)}]{vigil89}%
  \BibitemOpen
  \bibfield  {author} {\bibinfo {author} {\bibfnamefont {R.~S.}\ \bibnamefont
  {Vigil}}\ and\ \bibinfo {author} {\bibfnamefont {R.~M.}\ \bibnamefont
  {Ziff}},\ }\bibfield  {title} {\enquote {\bibinfo {title} {Random sequential
  adsorption of unoriented rectangles onto a plane},}\ }\href@noop {}
  {\bibfield  {journal} {\bibinfo  {journal} {J. Chem. Phys.}\ }\textbf
  {\bibinfo {volume} {91}},\ \bibinfo {pages} {2599} (\bibinfo {year}
  {1989})}\BibitemShut {NoStop}%
\bibitem [{\citenamefont {Feder}(1980)}]{feder80}%
  \BibitemOpen
  \bibfield  {author} {\bibinfo {author} {\bibfnamefont {J.}~\bibnamefont
  {Feder}},\ }\bibfield  {title} {\enquote {\bibinfo {title} {Random sequential
  adsorption},}\ }\href@noop {} {\bibfield  {journal} {\bibinfo  {journal} {J.
  Theor. Bio.}\ }\textbf {\bibinfo {volume} {87}},\ \bibinfo {pages} {237}
  (\bibinfo {year} {1980})}\BibitemShut {NoStop}%
\bibitem [{\citenamefont {Zhang}\ and\ \citenamefont
  {Torquato}(2013)}]{zhang13}%
  \BibitemOpen
  \bibfield  {author} {\bibinfo {author} {\bibfnamefont {G.}~\bibnamefont
  {Zhang}}\ and\ \bibinfo {author} {\bibfnamefont {S.}~\bibnamefont
  {Torquato}},\ }\bibfield  {title} {\enquote {\bibinfo {title} {Precise
  algorithm to generate random sequential addition of hard hyperspheres at
  saturation},}\ }\href@noop {} {\bibfield  {journal} {\bibinfo  {journal}
  {Phys. Rev. E}\ }\textbf {\bibinfo {volume} {88}},\ \bibinfo {pages} {053312}
  (\bibinfo {year} {2013})}\BibitemShut {NoStop}%
\bibitem [{\citenamefont {Onsager}(1949)}]{onsager49}%
  \BibitemOpen
  \bibfield  {author} {\bibinfo {author} {\bibfnamefont {L.}~\bibnamefont
  {Onsager}},\ }\bibfield  {title} {\enquote {\bibinfo {title} {The effects of
  shape on the interaction of colloidal particles},}\ }\href@noop {} {\bibfield
   {journal} {\bibinfo  {journal} {Ann. New York Acad. Sci.}\ }\textbf
  {\bibinfo {volume} {51}},\ \bibinfo {pages} {627} (\bibinfo {year}
  {1949})}\BibitemShut {NoStop}%
\bibitem [{\citenamefont {Ricci}\ \emph {et~al.}(1994)\citenamefont {Ricci},
  \citenamefont {Talbot}, \citenamefont {Tarjus},\ and\ \citenamefont
  {Viot}}]{ricci94}%
  \BibitemOpen
  \bibfield  {author} {\bibinfo {author} {\bibfnamefont {S.~M.}\ \bibnamefont
  {Ricci}}, \bibinfo {author} {\bibfnamefont {J.}~\bibnamefont {Talbot}},
  \bibinfo {author} {\bibfnamefont {G.}~\bibnamefont {Tarjus}}, \ and\ \bibinfo
  {author} {\bibfnamefont {P.}~\bibnamefont {Viot}},\ }\bibfield  {title}
  {\enquote {\bibinfo {title} {A structural comparison of random sequential
  adsorption and equilibrium configurations of spherocylinders},}\ }\href@noop
  {} {\bibfield  {journal} {\bibinfo  {journal} {J. Chem. Phys.}\ }\textbf
  {\bibinfo {volume} {101}},\ \bibinfo {pages} {9164} (\bibinfo {year}
  {1994})}\BibitemShut {NoStop}%
\bibitem [{\citenamefont {de~Gennes}(1979)}]{degennes79}%
  \BibitemOpen
  \bibfield  {author} {\bibinfo {author} {\bibfnamefont {P.~G.}\ \bibnamefont
  {de~Gennes}},\ }\href@noop {} {\emph {\bibinfo {title} {Scaling Concepts in
  Polymer Physics}}}\ (\bibinfo  {publisher} {Cornell University Press
  (Ithaca)},\ \bibinfo {year} {1979})\BibitemShut {NoStop}%
\bibitem [{foo({\natexlab{a}})}]{footSofQvec}%
  \BibitemOpen
  \href@noop {} \bibinfo {note} {$S(\vec{q})$ can in
  principle depend on the direction of $\vec{q}$, but it is isotropic for all
  the systems studied in this paper because none of them possess long-range
  orientational order. One could try to define a $S(q, \Delta\theta)$ that is
  analogous to $g(r, \Delta\theta)$, but such a quantity does not have an
  obvious geometrical interpretation.}\BibitemShut {Stop}%
\bibitem [{\citenamefont {Xu}\ \emph {et~al.}(2013)\citenamefont {Xu},
  \citenamefont {Li}, \citenamefont {Sun},\ and\ \citenamefont {An}}]{xu13b}%
  \BibitemOpen
  \bibfield  {author} {\bibinfo {author} {\bibfnamefont {W.-S.}\ \bibnamefont
  {Xu}}, \bibinfo {author} {\bibfnamefont {Y.-W.}\ \bibnamefont {Li}}, \bibinfo
  {author} {\bibfnamefont {Z.-Y.}\ \bibnamefont {Sun}}, \ and\ \bibinfo
  {author} {\bibfnamefont {L.-J.}\ \bibnamefont {An}},\ }\bibfield  {title}
  {\enquote {\bibinfo {title} {Hard ellipses: Equation of state, structure, and
  self-diffusion},}\ }\href@noop {} {\bibfield  {journal} {\bibinfo  {journal}
  {J. Chem. Phys.}\ }\textbf {\bibinfo {volume} {139}},\ \bibinfo {pages}
  {024501} (\bibinfo {year} {2013})}\BibitemShut {NoStop}%
\bibitem [{\citenamefont {Zheng}\ \emph {et~al.}(2021)\citenamefont {Zheng},
  \citenamefont {Ni}, \citenamefont {Wang},\ and\ \citenamefont
  {Han}}]{zheng21}%
  \BibitemOpen
  \bibfield  {author} {\bibinfo {author} {\bibfnamefont {Z.}~\bibnamefont
  {Zheng}}, \bibinfo {author} {\bibfnamefont {R.}~\bibnamefont {Ni}}, \bibinfo
  {author} {\bibfnamefont {Y.}~\bibnamefont {Wang}}, \ and\ \bibinfo {author}
  {\bibfnamefont {Y.}~\bibnamefont {Han}},\ }\bibfield  {title} {\enquote
  {\bibinfo {title} {Translational and rotational critical-like behaviors in
  the glass transition of colloidal ellipsoid monolayers},}\ }\href@noop {}
  {\bibfield  {journal} {\bibinfo  {journal} {Sci. Adv.}\ }\textbf {\bibinfo
  {volume} {7}},\ \bibinfo {pages} {eabd1958} (\bibinfo {year}
  {2021})}\BibitemShut {NoStop}%
\bibitem [{\citenamefont {Tan}\ \emph {et~al.}(2021)\citenamefont {Tan},
  \citenamefont {Chen}, \citenamefont {Wang}, \citenamefont {Zhang},\ and\
  \citenamefont {Ling}}]{tan21}%
  \BibitemOpen
  \bibfield  {author} {\bibinfo {author} {\bibfnamefont {X.}~\bibnamefont
  {Tan}}, \bibinfo {author} {\bibfnamefont {U.}~\bibnamefont {Chen}}, \bibinfo
  {author} {\bibfnamefont {H.}~\bibnamefont {Wang}}, \bibinfo {author}
  {\bibfnamefont {Z.}~\bibnamefont {Zhang}}, \ and\ \bibinfo {author}
  {\bibfnamefont {X.~S.}\ \bibnamefont {Ling}},\ }\bibfield  {title} {\enquote
  {\bibinfo {title} {2d isotropic-nematic transition in colloidal suspensions
  of ellipsoids},}\ }\href@noop {} {\bibfield  {journal} {\bibinfo  {journal}
  {Soft Matt.}\ }\textbf {\bibinfo {volume} {17}},\ \bibinfo {pages} {6001}
  (\bibinfo {year} {2021})}\BibitemShut {NoStop}%
\bibitem [{\citenamefont {Wang}\ \emph {et~al.}(2018)\citenamefont {Wang},
  \citenamefont {Mei}, \citenamefont {Song}, \citenamefont {Lu},\ and\
  \citenamefont {An}}]{wang18b}%
  \BibitemOpen
  \bibfield  {author} {\bibinfo {author} {\bibfnamefont {L.}~\bibnamefont
  {Wang}}, \bibinfo {author} {\bibfnamefont {B.}~\bibnamefont {Mei}}, \bibinfo
  {author} {\bibfnamefont {J.}~\bibnamefont {Song}}, \bibinfo {author}
  {\bibfnamefont {Y.}~\bibnamefont {Lu}}, \ and\ \bibinfo {author}
  {\bibfnamefont {L.}~\bibnamefont {An}},\ }\bibfield  {title} {\enquote
  {\bibinfo {title} {Structural relaxation and glass transition behavior of
  binary hard-ellipse mixtures},}\ }\href@noop {} {\bibfield  {journal}
  {\bibinfo  {journal} {Sci. China Chem.}\ }\textbf {\bibinfo {volume} {61}},\
  \bibinfo {pages} {613} (\bibinfo {year} {2018})}\BibitemShut {NoStop}%
\bibitem [{\citenamefont {Bautista-Carbajal}\ and\ \citenamefont
  {Odriozola}(2014)}]{bautista14b}%
  \BibitemOpen
  \bibfield  {author} {\bibinfo {author} {\bibfnamefont {G.}~\bibnamefont
  {Bautista-Carbajal}}\ and\ \bibinfo {author} {\bibfnamefont {G.}~\bibnamefont
  {Odriozola}},\ }\bibfield  {title} {\enquote {\bibinfo {title} {Phase diagram
  of two-dimensional hard ellipses},}\ }\href@noop {} {\bibfield  {journal}
  {\bibinfo  {journal} {J. Chem. Phys.}\ }\textbf {\bibinfo {volume} {140}},\
  \bibinfo {pages} {204502} (\bibinfo {year} {2014})}\BibitemShut {NoStop}%
\bibitem [{\citenamefont {Ziff}\ and\ \citenamefont {Vigil}(1990)}]{ziff90}%
  \BibitemOpen
  \bibfield  {author} {\bibinfo {author} {\bibfnamefont {R.~M.}\ \bibnamefont
  {Ziff}}\ and\ \bibinfo {author} {\bibfnamefont {R.~D.}\ \bibnamefont
  {Vigil}},\ }\bibfield  {title} {\enquote {\bibinfo {title} {Kinetics and
  fractal properties of the random sequential adsorption of line segments},}\
  }\href@noop {} {\bibfield  {journal} {\bibinfo  {journal} {J. Phys. A: Math.
  Gen.}\ }\textbf {\bibinfo {volume} {23}},\ \bibinfo {pages} {5103} (\bibinfo
  {year} {1990})}\BibitemShut {NoStop}%
\bibitem [{foo({\natexlab{b}})}]{footTStrans}%
  \BibitemOpen
  \href@noop {}  \bibinfo {note} {A comparable transition
  for spherocylinders was reported in Ref.\ \cite{ricci94}, but it was not examined
  in detail.}\BibitemShut {Stop}%
\bibitem [{\citenamefont {Chong}\ and\ \citenamefont {Kob}(2009)}]{chong09}%
  \BibitemOpen
  \bibfield  {author} {\bibinfo {author} {\bibfnamefont {S.-H.}\ \bibnamefont
  {Chong}}\ and\ \bibinfo {author} {\bibfnamefont {W.}~\bibnamefont {Kob}},\
  }\bibfield  {title} {\enquote {\bibinfo {title} {Coupling and decoupling
  between translational and rotational dynamics in a supercooled molecular
  liquid},}\ }\href@noop {} {\bibfield  {journal} {\bibinfo  {journal} {Phys.
  Rev. Lett.}\ }\textbf {\bibinfo {volume} {102}},\ \bibinfo {pages} {025702}
  (\bibinfo {year} {2009})}\BibitemShut {NoStop}%
\bibitem [{\citenamefont {Mishra}\ \emph {et~al.}(2014)\citenamefont {Mishra},
  \citenamefont {Nagamanasa}, \citenamefont {Ganapathy}, \citenamefont {Sood},\
  and\ \citenamefont {Gokhale}}]{mishra14}%
  \BibitemOpen
  \bibfield  {author} {\bibinfo {author} {\bibfnamefont {C.~K.}\ \bibnamefont
  {Mishra}}, \bibinfo {author} {\bibfnamefont {K.~H.}\ \bibnamefont
  {Nagamanasa}}, \bibinfo {author} {\bibfnamefont {R.}~\bibnamefont
  {Ganapathy}}, \bibinfo {author} {\bibfnamefont {A.~K.}\ \bibnamefont {Sood}},
  \ and\ \bibinfo {author} {\bibfnamefont {S.}~\bibnamefont {Gokhale}},\
  }\bibfield  {title} {\enquote {\bibinfo {title} {Dynamical facilitation
  governs glassy dynamics in suspensions of colloidal ellipsoids},}\
  }\href@noop {} {\bibfield  {journal} {\bibinfo  {journal} {Proc. Nat. Acad.
  Sci.}\ }\textbf {\bibinfo {volume} {111}},\ \bibinfo {pages} {15362}
  (\bibinfo {year} {2014})}\BibitemShut {NoStop}%
\bibitem [{\citenamefont {Mishra}\ and\ \citenamefont
  {Ganapathy}(2015)}]{mishra15}%
  \BibitemOpen
  \bibfield  {author} {\bibinfo {author} {\bibfnamefont {C.~K.}\ \bibnamefont
  {Mishra}}\ and\ \bibinfo {author} {\bibfnamefont {R.}~\bibnamefont
  {Ganapathy}},\ }\bibfield  {title} {\enquote {\bibinfo {title} {Shape of
  dynamical heterogeneities and fractional stokes-einstein and
  stokes-einstein-debye relations in quasi-two-dimensional suspensions of
  colloidal ellipsoids},}\ }\href@noop {} {\bibfield  {journal} {\bibinfo
  {journal} {Phys. Rev. Lett.}\ }\textbf {\bibinfo {volume} {114}},\ \bibinfo
  {pages} {198302} (\bibinfo {year} {2015})}\BibitemShut {NoStop}%
\bibitem [{\citenamefont {Roller}\ \emph {et~al.}(2020)\citenamefont {Roller},
  \citenamefont {Geiger}, \citenamefont {Voggenreiter}, \citenamefont
  {Meijer},\ and\ \citenamefont {Zumbusch}}]{roller20}%
  \BibitemOpen
  \bibfield  {author} {\bibinfo {author} {\bibfnamefont {J.}~\bibnamefont
  {Roller}}, \bibinfo {author} {\bibfnamefont {J.~D.}\ \bibnamefont {Geiger}},
  \bibinfo {author} {\bibfnamefont {M.}~\bibnamefont {Voggenreiter}}, \bibinfo
  {author} {\bibfnamefont {J.-M.}\ \bibnamefont {Meijer}}, \ and\ \bibinfo
  {author} {\bibfnamefont {A.}~\bibnamefont {Zumbusch}},\ }\bibfield  {title}
  {\enquote {\bibinfo {title} {Formation of nematic order in 3d systems of hard
  colloidal ellipsoids},}\ }\href@noop {} {\bibfield  {journal} {\bibinfo
  {journal} {Soft Matt.}\ }\textbf {\bibinfo {volume} {16}},\ \bibinfo {pages}
  {1021} (\bibinfo {year} {2020})}\BibitemShut {NoStop}%
\bibitem [{\citenamefont {Pal}\ \emph {et~al.}(2020)\citenamefont {Pal},
  \citenamefont {Martinez}, \citenamefont {Ito}, \citenamefont {Arlt},
  \citenamefont {Crassous}, \citenamefont {Poon},\ and\ \citenamefont
  {Schurtenberger}}]{pal20}%
  \BibitemOpen
  \bibfield  {author} {\bibinfo {author} {\bibfnamefont {A.}~\bibnamefont
  {Pal}}, \bibinfo {author} {\bibfnamefont {V.~A.}\ \bibnamefont {Martinez}},
  \bibinfo {author} {\bibfnamefont {T.~H.}\ \bibnamefont {Ito}}, \bibinfo
  {author} {\bibfnamefont {J.}~\bibnamefont {Arlt}}, \bibinfo {author}
  {\bibfnamefont {J.~J.}\ \bibnamefont {Crassous}}, \bibinfo {author}
  {\bibfnamefont {W.~C.~K.}\ \bibnamefont {Poon}}, \ and\ \bibinfo {author}
  {\bibfnamefont {P.}~\bibnamefont {Schurtenberger}},\ }\bibfield  {title}
  {\enquote {\bibinfo {title} {Anisotropic dynamics and kinetic arrest of dense
  colloidal ellipsoids in the presence of an external field studied by
  differential dynamic microscopy},}\ }\href@noop {} {\bibfield  {journal}
  {\bibinfo  {journal} {Sci. Adv.}\ }\textbf {\bibinfo {volume} {6}},\ \bibinfo
  {pages} {eaaw9733} (\bibinfo {year} {2020})}\BibitemShut {NoStop}%
\bibitem [{\citenamefont {Pal}\ \emph {et~al.}(2021)\citenamefont {Pal},
  \citenamefont {Kamal}, \citenamefont {Holmqvist},\ and\ \citenamefont
  {Schurtenberger}}]{pal21}%
  \BibitemOpen
  \bibfield  {author} {\bibinfo {author} {\bibfnamefont {A.}~\bibnamefont
  {Pal}}, \bibinfo {author} {\bibfnamefont {M.~A.}\ \bibnamefont {Kamal}},
  \bibinfo {author} {\bibfnamefont {P.}~\bibnamefont {Holmqvist}}, \ and\
  \bibinfo {author} {\bibfnamefont {P.}~\bibnamefont {Schurtenberger}},\
  }\bibfield  {title} {\enquote {\bibinfo {title} {Structure and dynamics of
  dense colloidal ellipsoids at the nearest-neighbor length scale},}\
  }\href@noop {} {\bibfield  {journal} {\bibinfo  {journal} {Phys. Rev.
  Research}\ }\textbf {\bibinfo {volume} {3}},\ \bibinfo {pages} {023254}
  (\bibinfo {year} {2021})}\BibitemShut {NoStop}%
\bibitem [{\citenamefont {Jadrich}\ \emph {et~al.}(2022)\citenamefont
  {Jadrich}, \citenamefont {Lundquist},\ and\ \citenamefont
  {Truskett}}]{jadrich22}%
  \BibitemOpen
  \bibfield  {author} {\bibinfo {author} {\bibfnamefont {R.~B.}\ \bibnamefont
  {Jadrich}}, \bibinfo {author} {\bibfnamefont {B.~A.}\ \bibnamefont
  {Lundquist}}, \ and\ \bibinfo {author} {\bibfnamefont {T.~M.}\ \bibnamefont
  {Truskett}},\ }\bibfield  {title} {\enquote {\bibinfo {title} {Treating
  random sequential addition via the replica method},}\ }\href@noop {}
  {\bibfield  {journal} {\bibinfo  {journal} {J. Chem. Phys.}\ }\textbf
  {\bibinfo {volume} {157}} (\bibinfo {year} {2022})}\BibitemShut {NoStop}%
\end{thebibliography}

%

\end{document}